\documentclass[aps,prl,a4paper,reprint,showpacs,amsmath,amssymb,superscriptaddress,floatfix]{revtex4-1}

\usepackage{amssymb}
\usepackage{latexsym}
\usepackage[dvips]{graphicx}
\usepackage{epsfig}

\usepackage{dcolumn}
\usepackage{amsmath}
\usepackage{amsfonts}
\usepackage{bm}
\usepackage{color}

\newcommand{\be}{\begin{equation}}
\newcommand{\ee}{\end{equation}}
\newcommand{\bea}{\begin{eqnarray}}
\newcommand{\eea}{\end{eqnarray}}

\newcommand{\p}{\partial}
\newcommand{\s}{\sigma}

\newcommand{\la}{\langle}
\newcommand{\ra}{\rangle}
\newcommand{\rd}{\mbox{d}}

\newcommand{\ri}{\mbox{i}}
\newcommand{\re}{\mbox{e}}

\renewcommand{\vec}[1]{{\bm #1}}

\begin{document}
\title{Superconductor-Metal Transition in Odd-Frequency Paired Superconductor in Magnetic Field}
\author{A. M. Tsvelik}
\affiliation{Condensed Matter Physics and Materials Science Division, \\
Brookhaven National Laboratory, Upton, NY 11973-5000, USA}


\date{\today }

\begin{abstract}
It is shown that the application of sufficiently strong magnetic field to the odd-frequency paired Pair Density Wave state described in Phys. Rev. B{\bf 94}, 165114 (2016) leads to formation of a low temperature metallic state with zero Hall response.  Applications of these ideas to the recent experiments on stripe-ordered La$_{1.875}$Ba$_{0.125}$CuO$_4$ are discussed.  
\end{abstract}


\maketitle

{\bf Significance Statement. } It is generally expected that when magnetic field destroys superconductivity in two dimensions the system becomes an insulator. It is shown that there is a type of superconductivity, namely the one where the wave function of pairs is odd in time, where the result is not an insulator, but a metal with a zero Hall response. It is suggested that the transition recently observed in the striped-ordered high $T_c$ superconductor La$_{1.875}$Ba$_{0.125}$CuO$_4$ may belong to this category.

{\it Introduction}. Recent magnetotransport measurements in x=1/8 LBCO \cite{tranquada} have revealed yet additional extraordinary features of this otherwise highly usual system. It has turned out that when the applied magnetic field destroys the superconductivity in this layered material, the system becomes metallic with zero Hall response. This behavior is robust down to the lowest temperatures; the sheet resistance gradually increases with magnetic field leveling off at around $B \sim 30$T at $G = 2e^2/h$. 

 At x=1/8 doping the holes in copper oxide layers of LBCO are arranged in static stripes at temperatures below $\sim 40$K. The material  undergoes Berezinskii-Kosterlitz-Thouless (BKT) transition at around $T_{BKT} = 16$K into a two dimensional superconducting phase with a finite resistivity in the $c$-direction \cite{BKT}. The Meissner effect is established at much lower temperature $\sim 3$K. The theoretical explanation put forward in \cite{berg} assign these unusual properties to the formation in each CuO layer of Pair Density Wave (PDW) - a superconducting state where the pairs have nonzero momentum ${\bf Q}$. If the direction of ${\bf Q}$ is different in neighboring copper oxide layers than the pairs would not able to tunnel and the layers would remain decoupled. The theory \cite{berg,FKT} models the PDW state as an array of doped chains separated by undoped regions; the chains contain Luther-Emery liquids with gapped spin sector and enhanced superconducting fluctuations. An isolated chain has a quasi long range superconducting order with a spin gap; the chains interact through Josephson coupling (pair tunneling) and the long range Coulomb interaction. Quasiparticles play no active role in this scenario. 
 I argue  that this standard picture of the stripe phase cannot explain the transport data of \cite{tranquada}, namely the combination of metallic longitudinal resistivity and zero Hall conductivity. It will be shown that once the strong magnetic field makes the Josephson coupling irrelevant, the superconducting correlations in the transverse direction  become short range suppressing  the transport in the direction perpendicular to the chains. To explain the metallic transport one has to assume the presence of quasiparticles, as was done in \cite{orgad}. However, then one has to explain zero Hall conductivity.  
  
  The present paper suggests a different scenario in which the above difficulties are resolved. It is based on the results obtained in \cite{tsvelik2016}. This paper describes a version of a striped model where the spin gap and superconducting coupling on the hole doped stripes come as a result of exchange interactions between the holes and the surrounding spins and the Heisenberg exchange between the spins. This leads to formation of PDW with the wave vector along the stripes together with formation of hole- and electron like Fermi pockets of gapless quasiparticles. The restrictions  related to the Luttinger theorem guarantee the equality of number of electrons and holes and, as a consequence, zero Hall response. The existence of ungapped  quasiparticles is due to  the fact that the PDW order parameter had a finite wave vector incommensurate with the Fermi surface which  eliminates coupling between the order parameter and the  quasiparticles. The superconductivity is essentially 2D as in the standard layered model, but since  nothing prevents quasiparticles from tunneling between the layers, and their transport  is 3D.

  
  The present paper begins with a pedagogical description of this model adopted to the case of a layered 3D material with stripes in neighboring layers running perpendicular to each other, as is the case with LBCO.  The model displays  staggered odd-frequency paired PDW with a wave vector directed along the stripes. This staggering makes the interlayer coupling of the order parameters difficult. Below I will recall the main results of \cite{tsvelik2016} generalizing them for finite magnetic fields and putting them in the context of \cite{tranquada}.
  
   


 {\it The model.} The adopted description of the striped state is one of a Kondo lattice. This is, however, a lattice of a special kind where the conduction electrons and local moments are segregated into stripes. In the first approximation we can consider a two dimensional arrangement of parallel stripes. The salient feature of the model  is incommensurability between the Fermi wave vector of the holes occupying the conducting chains (stripes) and the lattice. The standard thinking about 
 Kondo lattices 
 considers its physics as a product of competition between the inter-spin and the Kondo  interactions. If the former one wins the spins decouple from the electrons and when the latter wins the spins fractionalize, hybridize with the itinerant electrons and become a part of the conduction band giving rise to a heavy fermion Fermi liquid. 
 It has been frequently suggested (see, for example, \cite{subir, piers}) that there are  circumstances when the spin spin system left to its own devices will not magnetically order, but form a liquid - a strongly correlated state with short range spin correlations. However, the experience of  many years of research in this direction indicates that such disordered states are very difficult to realize. If interacting spins do not order magnetically they tend to form so-called Valence Bond Solids where the magnetic excitations are gapped, but the translational symmetry is still broken. 
 
  In \cite{tsvelik2016} I have considered a mechanism of spin liquid formation  based on {\it cooperation} between the Kondo and the Heisenberg exchange interactions. Somewhat paradoxically such cooperation works better when the Heisenberg exchange is stronger than the Kondo one provided the spin liquid has a Fermi surface as is the case for solitary spin S=1/2 Heisenberg chains. Such situation takes place already for a single spin chain and, indeed, a single Heisenberg chain coupled to one-dimensional electron gas (1DEG) already provides a mechanism for Pair Density Wave formation, as has been noticed in \cite{zachar, KHKivelson}. Hence the simplest way to realize such situation is to consider an array of spin S=1/2 Heisenberg chains decoupled from each other. In LBCO the doped and undoped stripes alternate. To simplify matters I consider a somewhat different  situation when doped chains lie on top of the  Heisenberg one as was done in \cite{tsvelik2016}. In this case each  spin chain is coupled to only one conducting chain.  This arrangement allows me not to consider additional details which would only muddle the discussion. 
  


\begin{figure}[htp]
\centerline{\includegraphics[angle = 0,
width=0.5\columnwidth]{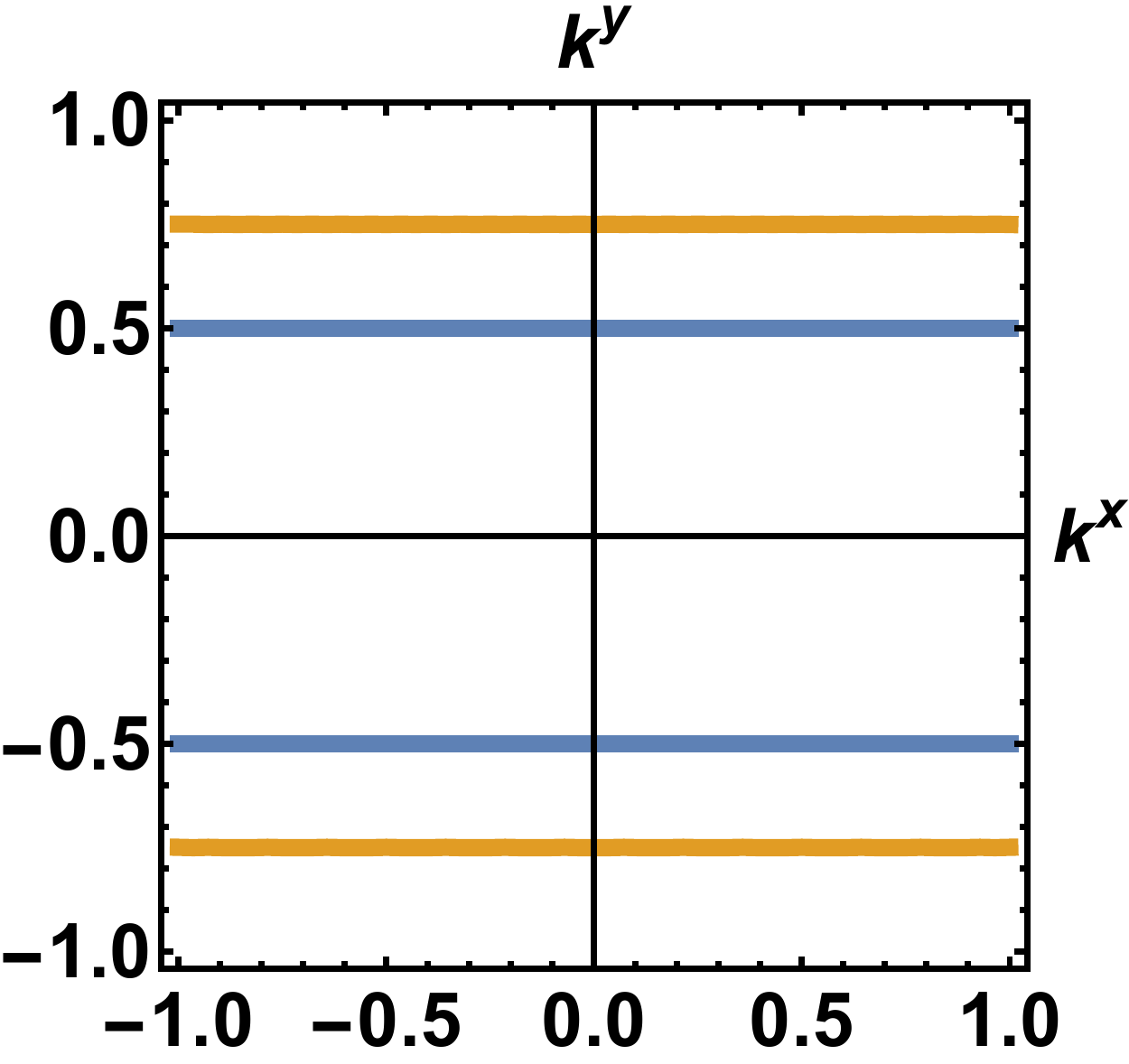}}
\centerline{\includegraphics[angle = 0,
width=0.5\columnwidth]{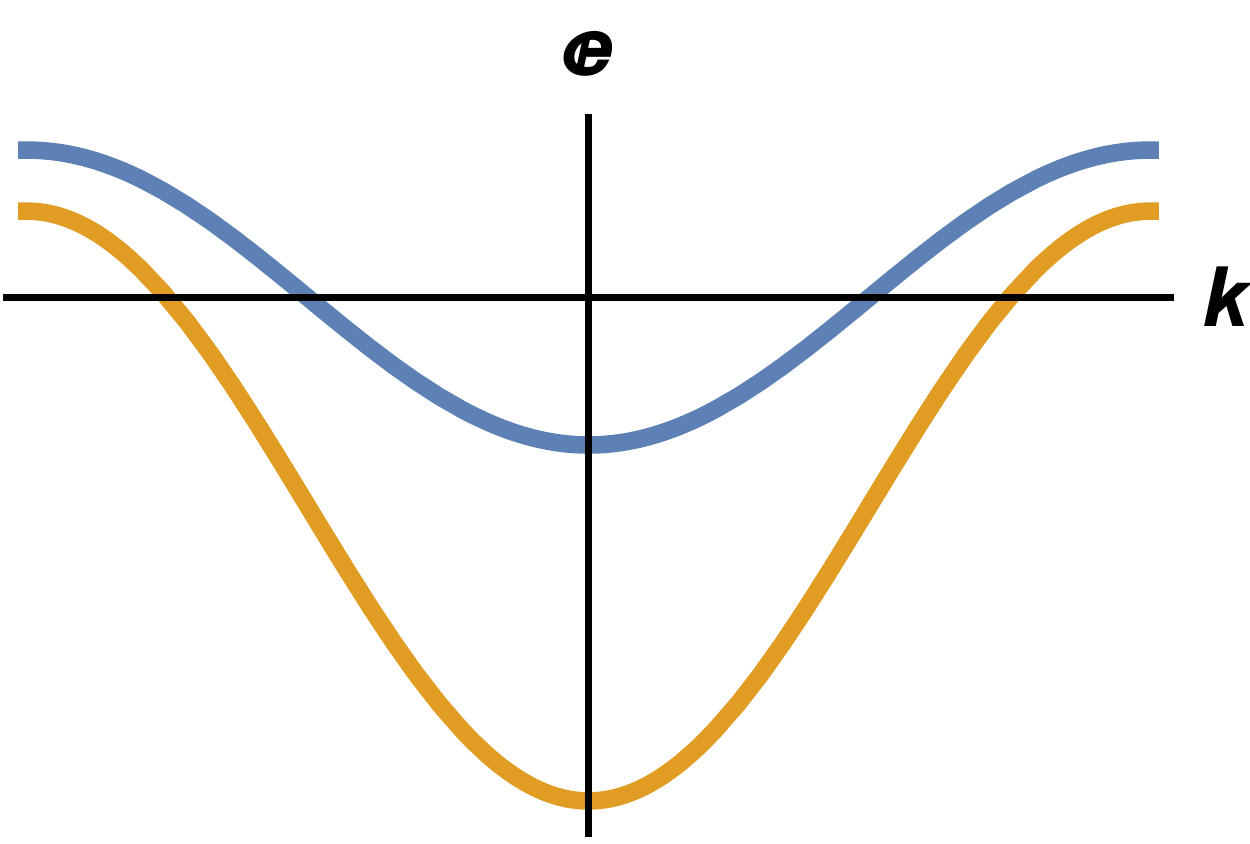}}
\caption{ A.) Spinon (blue) and electron (yellow) Fermi surfaces  in the array of 1D Kondo-Heisenberg ladders. In the limit when ladders are decoupled the Fermi surfaces are flat and exhibit a perfect nesting. Then spinons and electrons with opposite chirality hybridize and create spin gaps. The spin subsystem decouples into two independent spin liquids made as indicated by arrows. B.)  The bare spinon (blue) and electron (yellow) spectra. The Fermi momenta do not coincide. }
\label{Fig3}
\end{figure}
 A single  KH ladder consists  of  an antiferromagnetic spin S=1/2 Heisenberg chain (HC) coupled to 1DEG via an anferromagnetic exchange interaction:
\bea
&& H = \sum_k \epsilon(k) \psi^+_{k\s}\psi_{k\s} + \frac{J_K}{2} \sum_{k,q}\psi^+_{k+q,\alpha}\vec\s_{\alpha\beta}\psi_{k,\beta} {\bf S}_q + \nonumber\\
&& J_H\sum_n {\bf S}_n{\bf S}_{n+1}, \label{model1}
\eea
where $\psi^+,\psi$ are creation and annihilation operators of the 1DEG, $\s^a$ are the Pauli matrices, ${\bf S}_n$ is the spin S=1/2 operator on site $n$ and ${\bf S}_q$ is its Fourier transform.   It is assumed that $J_K << J_H$ and  the 1DEG is far from half filling, $|2k_Fa_0 -\pi| \sim 1$ ($k_F$ is the Fermi momentum of the electrons). Under these assumptions one can formulate the low energy description of (\ref{model1}), taking into account that the backscattering processes between excitations in the HC and the 1DEG are suppressed by the incommensurability of the 1DEG.  The effective theory is valid for energies much smaller than both the Fermi energy $\epsilon_F$ and the Heisenberg exchange interaction  $J_H$ of the model (\ref{model1}). It is  integrable \cite{zachar} and  the exact solution  was used as a springboard for a controllable approach to the model  of a D-dimensional  array of   KH ladders developed in \cite{tsvelik2016}.

At $J_K =0$ both 1DEG and the Heisenberg chain are critical systems. The excitations of the HC are gapless spinons whose spectrum is linear at small momenta: $\omega = v_H|k|$ with $v_H = \pi J_H/2$. Spinons are fractionalized particles: they carry zero electric charge and spin S=1/2.
In the absence of umklapp the only smooth parts of the magnetizations of spin and electron chains couple. It is remarkable that in the spin S=1/2 Heisenberg chain the smooth part can be represented as a sum of  the spin currents ${\bf j}_{R,L}$ (\ref{S}) which allow a fermionic representation: $
{\bf j}_{R}=\frac{1}{2}r^+\vec\s r, ~~{\bf j}_L = \frac{1}{2}l^+\vec\s l, $
 where $r,l$ are noninteracting 1D fermions with dispersion $\pm v_H k_x$. These fermions carry fictitious U(1) charge. However, the charge degrees of freedom do not affect  the current-current commutation relations and hence do not partake in the interaction with the conduction electrons. When the Kondo coupling is much smaller than the electron band width we can linearize the electron spectrum close to the Fermi points introducing right- and left moving fermions $R({\bf k}) = \psi(k_x+k_F,k_y), ~~L({\bf k}) = \psi(k_x-k_F,k_y)$. The resulting low energy description is 
\bea
&& H = H_+ + H_-\\
 &&H_+ = \sum_{\bf k}\Big\{ \epsilon_R({\bf k})R^+_{\alpha}({\bf k})R_{\alpha}({\bf k}) + v_Hk_x l^+_{\alpha}({\bf k})l_{\alpha}({\bf k}) \Big\} +\nonumber\\
 && J_K\int \rd V R^+\vec\s R(r) l^+\vec\s l(r) \label{big1}\\
 && H_- = H_+(R \rightarrow L, L \rightarrow R, k_x \rightarrow -k_x) \label{big2}.
 \eea
  We can choose $\epsilon_{R, L} = \pm v_F (k_x-k_F) + 2t(\cos k_y + \cos k_z)$.

   Representation (\ref{big1},\ref{big2}) is similar to the one frequently adopted for the Kondo lattices, see, for example, \cite{paschen1}. However, there is  one difference, namely that in our approach the spinon right- and left moving fermionic operators $r,l$ do not interact and in the standard treatment where the spins are arranged on a 3D lattice with isotropic interactions they do. 

 {\it The spectrum and the order parameters.}   The following simple mathematical description gives the gist of what is going on. Although it is possible to carry on the calculations rigorously, as was done in \cite{tsvelik2016}, I will resort to a simplified approach. Namely, I decouple the interaction with the Hubbard-Stratonovich transformation and look for the saddle point:
   \bea
   J_KR^+\vec\s R l^+\vec\s l \rightarrow  |\Delta_+|^2/2J_K + ( \Delta_+ R^+_{\alpha}l_{\alpha} + H.c.) , 
  \eea
  and the same for $L$ and $r$. Then the quasiparticle spectrum at the saddle point is
  \bea
 && E_{\pm}(k) = \pm (k_xv_H- \epsilon_R)/2 + \nonumber\\
 && \sqrt{(k_xv_H + \epsilon_R)^2/4 + |\Delta_+|^2}, \label{disp1}
\eea
where $\epsilon_R = v_Fk_x + t_{\perp}(k)$, where $t_{\perp}$ is the Fourier transform of the interchain tunneling . Strictly speaking, this procedure is justified when the SU(2) symmetry is replaced by the SU(N) one with $N>>1$. However, as was demonstrated in \cite{tsvelik2016}, the results remain robust even for $N=2$. Some details can be found in Supplementary Information \cite{SM}.   However, even on this level we see that $\Delta_{\pm}$ are complex fields and their phases must remain gapless.  

If electrons are also 1D their Fermi surface is flat and we have a process depicted on Fig. \ref{Fig3}A. It is assumed that the Fermi momenta of electrons and spinons are different so that there are no umklapp processes and the  hybridization takes place only between spinons and electrons of opposite chirality.  This opens a gap on the entire electron Fermi surface. If one allows an interchain tunneling the nesting becomes imperfect and pockets of electron- and hole-like quasiparticles will appear as on Fig. \ref{Fig4}. 

\begin{figure}
\centerline{\includegraphics[angle = 0,
width=0.4\columnwidth]{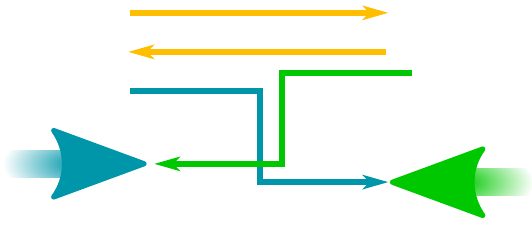}}
\centerline{\includegraphics[angle = 0,
width=0.4\columnwidth]{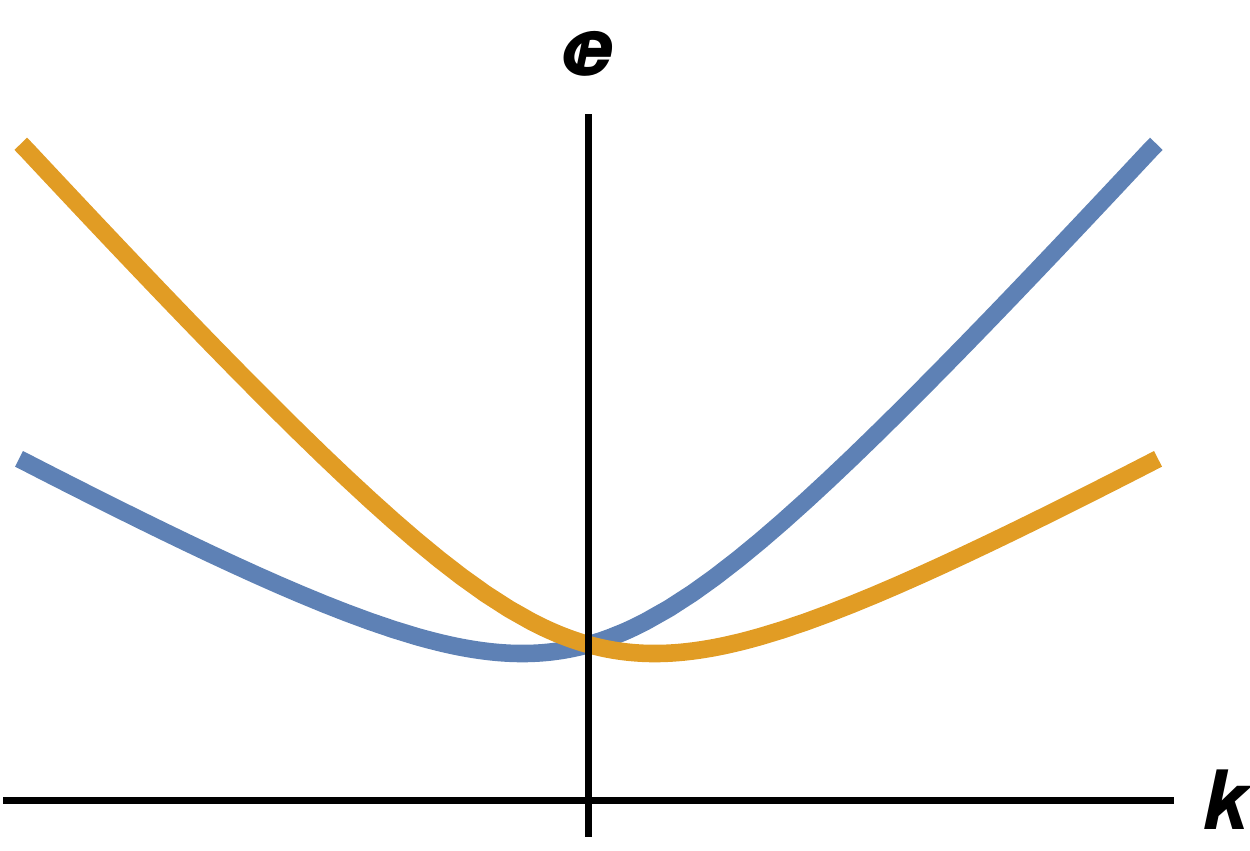}}
\caption{Holons of 1DEG (orange arrows) do not pair. Spinons of 1DEG (thin arrows) pair with the spinons of opposite chirality from the Heisenberg chain (thick arrows). This forms the gapped  spinon dispersion shown on the right. $e= E/\Delta$,  $q=k_x(v_Hv_F)^{1/2}/\Delta$ and $v_F/v_H =1/4$. The spinons of the Heisenberg model located at wave vectors $-\pi/2a_0$ (+$\pi/2a_0$) pair with the spinons of the 1DEG located at $k_F$ (-$k_F$). The product of the corresponding pairing amplitudes forms the amplitude $A$ of the composite order parameter (\ref{g}).}
 \label{Spectrum}
\end{figure}
The Hubbard-Stratonovich approach gives somewhat simplified picture of the spectrum. It turns out that the gapped parts of the spectrum (\ref{disp1}) corresponds to neutral spinons -  spin-1/2 incoherent excitations which remain confined to the chains. The gapless parts correspond to coherent quasiparticles whose Fermi surfaces in the form of particle and hole pockets are shown on Fig.\ref {Fig4}. 
\begin{figure}[!htp]
\centerline{\includegraphics[angle = 0,
width=0.6\columnwidth]
{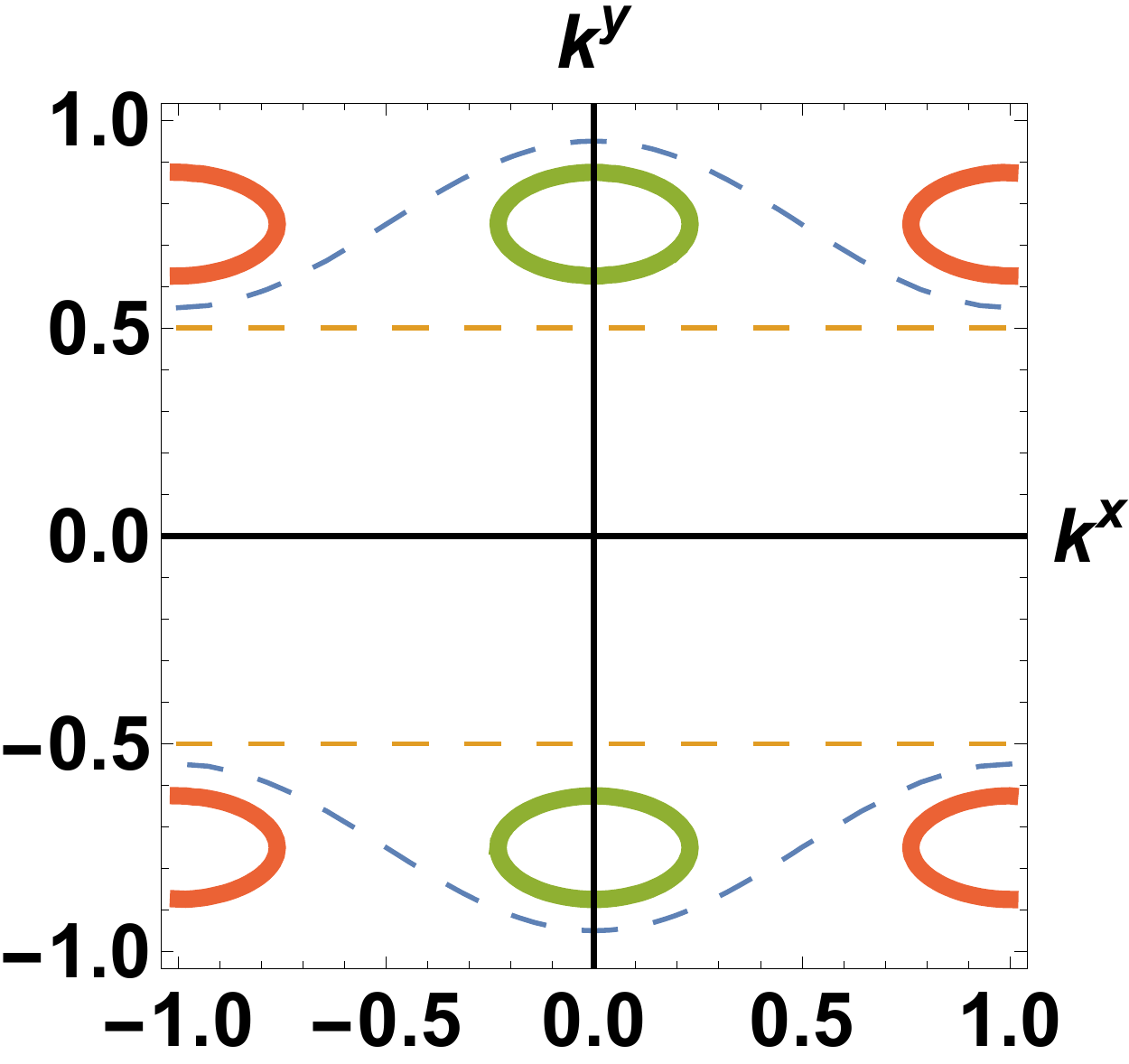}}
\caption{ Pockets of electron- and hole-like quasiparticles formed in the spin liquid state with a sufficiently strong  interstripe electron tunneling. The bare electron and spinon Fermi surfaces are gapped and marked by dashed lines. }
\label{Fig4}
\end{figure}
There are also gapless collective modes corresponding to fluctuations of the order parameter fields. The  Hubbard-Stratonovich fields $\Delta_{\pm}$ contain a fictitious U(1) phase of the $r,l$ fermions and hence are not gauge invariant. The real  order parameter (OP) fields in the KH ladder are their products $\Delta_+\Delta_-$ and $\Delta_+\Delta_-^*$\cite{footnote}.  
 These OPs can be expressed in terms of the electron and spin operators: 
\bea
&& {\cal O}_{cdw} =  \psi^+(x)\Big[({\bf S}_x{\bf S}_{x+a_0})\hat I + \ri(\vec\s {\bf S}_x)\Big]\psi(x)\re^{\ri(\pi/a_0 +2k_F)x}\nonumber\\
&& {\cal O}_{sc} =  \ri(-1)^{x/a_0}\psi(x)\s^y\Big[({\bf S}_x{\bf S}_{x+a_0})\hat I + \ri(\vec\s{\bf S}_x)\Big] \psi(x),\label{OP2}
\eea
where $\hat I$ is a unit matrix. For a single KH chain correlation functions of these  composite OPs  have a power law decay at $T=0$. These OPs can be conveniently written in the matrix form:
\bea
\hat {\cal O} = \left(
\begin{array}{cc}
{\cal O}_{cdw} & {\cal O}_{sc}^+\\
-{\cal O}_{sc} & {\cal O}_{cdw}^+\\
\end{array}
\right) =
 A\hat g, \label{g}
\eea
where $A\sim \Delta$ is an amplitude and $g$ is the matrix field of the SU$_1$(2) Wess-Zumino-Witten-Novikov model governing the dynamics of the collective charge excitations (see Eq.(\ref{charge} in \cite{SM}).

Using the equations of motion $\dot\psi = [H,\psi]$, where the dot stands for derivative in Matsubara time, one can show that these OPs have finite overlap with the order parameters of  the odd-frequency Pair Density Wave  and odd-frequency CDW:
\bea
&& {\cal O}_{osc} = \dot\psi(\tau, x)\s^y\psi(\tau,x) (-1)^{x/a_0}, \nonumber\\
&&  {\cal O}_{ocdw} = \dot\psi^+(\tau,x)\psi(\tau, x) \re^{\ri(2k_F + \pi/a_0)x}
\eea
One can find more detailed discussion of odd-frequency superconductivity in the review article \cite{balatsky}. The idea that Kondo lattices may support odd-frequency superconductivity was put forward in the 90-ties \cite{coleman1,coleman2} and its  relation to the composite orders (\ref{OP2}) was discussed in \cite{KHKivelson,georges,zachar}.  However, the mean field theory presented in \cite{coleman1,coleman2} was too simple to account for interesting properties of the KH ladder encoded in its correlation functions.

Let us now turn to the quasiparticles. The best way to detect them is to calculate the single particle Green's function. For the standard model such calculations were done in \cite{orgad} with the result that the PDW leaves certain parts of the Fermi surface ungapped. However, this approach does not produce  zero Hall response - one of the striking features on the metallic phase observed in \cite{tranquada}.  In the PDW state discuss here this feature comes as a consequence of the strong interactions and the spin gap formation. The simplest way to see this is to consider the  RPA form of the Green's function: 
\bea
G_{RPA} = [G^{-1}_{1D}(\omega,k_x) - t_{\perp}({\bf k})]^{-1}, \label{RPA}
\eea
where $G_{1D}$ is the the Green's function of a single KH ladder. This approximation allows one to take into account the strongest correlations on a single chain encoded in $G_{1D}$ which can be calculated nonperturbatively. Its  precise  form  is given in \cite{tsvelik2016,SM}. In the present context we need to know  that $G_{1D}(\omega=0, k_x = \pm k_{F}) =0$ which allows the purely one dimensional KH ladder to satisfy the Luttinger theorem despite the absence of Fermi surface. This property translates to (\ref{RPA}) which guaranties that even for sufficiently large $t_{\perp}$ when (\ref{RPA}) acquires quasiparticle poles at zero frequency, they will not contribute to the Luttinger volume already fixed by the zeroes. In other words, the poles will cancel each other resulting in  a compensated metal with zero Hall response \cite{SM}.

{\it A failure of the standard model.} Below I consider the standard model of stripes in strong magnetic field  and will show that ones the Josephson tunneling is frustrated the low temperature transport of pairs becomes impossible. The standard model describes the charge sector of stripe phase as an array of one dimensional Luther-Emery liquids coupled by Josephson tunneling. The effective Lagrangian density describing superconducting fluctuations of such system is 
\bea
&& {\cal L} = \sum_y \Big\{\frac{1}{2}\Big[v^{-1}(\p_{\tau}\theta_y)^2 +v(\p_x\theta_y)^2\Big] + \nonumber\\
&& J \cos[\beta(\theta_y - \theta_{y+1}) - 2hx/v]\Big\},
\eea
where $h = eHa_0v/c$ with $a_0$ being the interchain distance and $H$ the applied magnetic field; $v$ is the velocity of the phase mode,  $\beta$ is a  parameter related to the interactions. In what follows I set $v=1$. I assume that the long range Coulomb interaction is screened, for instance,  by the gapless quasiparticles as in \cite{orgad}.

 For our purposes it will be sufficient to calculate the order parameter correlation function 
 $\chi_P(\tau;x,y)=  \la\la \re^{\ri\beta\theta_y(\tau,x)}\re^{-\ri\beta\theta_0(0,0)}\ra\ra,$ 
for large magnetic field when it can be done using the perturbation theory. The expansion parameter of this theory is $J/h^{2-2d}$, where $d = \beta^2/4\pi$. In the leading order in this parameter the correlation function for a given $y$ must include only minimal number of Josephson interactions sufficient for the pair to tunnel for a given distance: 
\bea
&& \chi_P(A,B;y) = \label{F}\\
&& J^y\int \prod_{i=1}^y \rd\tau_i\rd x_i \chi(A,1)\chi^*(1,2)\chi(2,3)...\chi(y-1,B)\times\nonumber\\
&& \re^{2\ri h( x_1- x_2 + x_3 +...)},\nonumber
 \eea
where $A,B$ are shorthand for $(\tau_A,x_A), (\tau_B,x_B)$ and  $\chi(1,2)$ is just the correlation function $\chi_P(\tau_1-\tau_2, x_1-x_2;y=0)$  at $J=0$.  Taking the Fourier transform and setting $k_x=0$ to simplify the expressions, we get 
\bea
\chi_P(\omega,k_x=0,k_y) = \frac{G^{-1}(\omega,h) - J\cos k_y}{(G^{-1}(\omega,h) - J\cos k_y)^2 + J^2\sin^2k_y},\nonumber
\eea
where $G^{-1} = A(\omega^2 + h^2)^{1-d}$, with $A$ being nonuniversal dimensional  parameter. At $J << h^{2-2d}$ the correlator is short ranged which means that {|color{red} the transverse tunneling of pairs is blocked. At these circumstances one is left with conductivity along the chains, but since this is associated with charge density waves which are pinned by disorder, this will also vanish.} 

{\it Summary and Discussion}.  Let us summarize the physics of KHA  with the interstripe tunneling is of order of the spin gap.  At energies below the spin gap the system effectively splits into two quasi independent sectors. One sector is the collective modes - the superconducting and the CDW fluctuations. The other sector is the quasiparticles. The order parameters are staggered with wave vectors incommensurate with the Fermi surface which has the most profound consequences for the low temperature behavior. First of all the incommensurability  guarantees that the quasiparticle  Fermi surface remains ungapped even when there is a true long range order. Then in the layered system where the stripes in the neighboring planes are perpendicular to each other, the order becomes effectively two dimensional since the interlayer coupling is frustrated. The quasiparticle tunneling, however, is not frustrated and the quasiparticles are free to propagate in all directions which prevents their localization. At last, the total Fermi surface volume (the Fermi volume of the electron- minus the volume of the hole-like parts) is zero. As is explained above, this is a property of the strongly correlated spin liquid state from which the PDW originates. As a consequence, the Hall response is zero below the BKT transition. 

 Measurements of the specific heat produce a finite value of $\gamma(T \rightarrow 0) = 2.5$ mJ K$^{-2}$mol$^{-1}$ which increases to 2.8 mJ K$^{-2}$mol$^{-1}$ in H= 9T \cite{tranquada2008} (about an order of magnitude smaller than in the normal state). This is consistent with existence of a Fermi surface. The quasiparticle Fermi energy must be of the order of the spin gap which for a similar material LSCO is estimated as $\sim $9 meV \cite{tranquada2018}. Such shallow Fermi sea could remain  undetected by the ARPES measurements \cite{shen}  in the stripe-ordered  LBCO. In any case the ARPES experiments  represent problem for the standard model as well. Another potential problem is magnetic order in the stripe-ordered phase. Naturally, a strong order will destroy the spin gap which gives rise to the PDW. However, a weak order may coexist with PDW \cite{tsvelik2016,SM} and  the measurements in a similar compound LNSCO with $x=0.12$ yield a small Cu moment of 0.10 $\pm $0.03 $\mu_B$ \cite{tranquada96}. Besides zero Hall conductivity and finite $\gamma$ there is another feature which distinguishes the present theory from the standard one, it the direction of the wave vectors of the staggered order parameters. They are directed along the stripes and this presumably can be tested experimentally.

{\it Acknowledgements}. I am grateful to J. Tranquada for very useful discussions and to L. B. Ioffe and S. A. Kivelson to valuable remarks. The work was supported by Office of Basic Energy Sciences, Material Sciences and Engineering Division, U.S. Department of Energy (DOE) under Contract No. DE-SC0012704. 

\begin{widetext} 
\section{Supplementary Information} 

This material contains a technical description of the system described in the main text. This description partially repeats the presentation of \cite{Tsvelik2016}. 
\subsection{The core model: the Kondo-Heisenberg (KH) ladder}.


 The description given below is  based on the ideas of non-Abelian bosonization. The advantage of using this formalism is that it provides a uniform description for  the low energy Hamiltonians of 1DEG and spin S=1/2 Heisenberg model. Both Hamiltonians  can be expressed in terms of the current operators which form closed algebra, namely the SU$_1$(2) Kac-Moody one. This representation is extremely useful for nonperturbative treatment of the problem since from the very beginning it makes manifest several very nontrivial facts.

   The current  operators include the right- and the left moving components of the fermion fields which emerge from the low energy decomposition of the fermion field:
\be
 \psi(x) = \re^{-\ri k_F x} R(x) + \re^{\ri k_F x}L(x), \label{linear}
 \ee
 and smooth parts of the magnetization of the Heisenberg chain:
 \bea
{\bf S}_n = \Big[{\bf j}_R(x) + {\bf j}_L(x)\Big] + (-1)^n {\bf N}_s(x) +..., ~~ x= na_0\label{S}
\eea
where the dots stand for less relevant operators, $a_0$ is the lattice distance, ${\bf N}_s(x)$ is the staggered magnetization operator which will be discussed later in greater detail. As I have said, the spin currents ${\bf j}_R(x), {\bf j}_L(x)$  satisfy SU$_1$(2) Kac-Moody algebra as well as  the spin currents of the electrons
\be
{\bf F}_{R}=\frac{1}{2}R^+\vec\s R, ~~{\bf F}_L = \frac{1}{2}L^+\vec\s L,
 \ee
 and the electron charge currents:
 \be
 I^z_R = R^+_{\s}R_{\s}, ~~I^+_R= R^+_{\uparrow}R^+_{\downarrow}, ~~I^-_R = R_{\downarrow}R_{\uparrow}.
\ee

Due to the incommensurability between 1DEG and the lattice  the staggered magnetization drops out from the Kondo  interaction. As a result the latter interaction is expressed solely in terms of the currents. The resulting low energy Hamiltonian of the KH ladder is  \cite{Tsvelik2016} :
\bea
&& {\cal H}_{eff} = {\cal H}_{charge} 
+ {\cal H}_s^{(Rl)} + {\cal H}_s^{(Lr)} \label{model3}\\
&& {\cal H}_{charge} = \frac{2\pi v_F}{3}\Big( :{\bf I}_R{\bf I}_R: + :{\bf I}_L{\bf I}_L:\Big) \label{charge}\\
&& {\cal H}_s^{(Rl)} = \frac{2\pi v_F}{3}:{\bf F}_R{\bf F}_R: + \frac{2\pi v_H}{3}:{\bf j}_L{\bf j}_L: + J_K {\bf F}_R{\bf j}_L \label{GN1}\\
&& {\cal H}_s^{(Lr)} =  \frac{2\pi v_F}{3}:{\bf F}_L{\bf F}_L: + \frac{2\pi v_H}{3}:{\bf j}_R{\bf j}_R: + J_K {\bf F}_L{\bf j}_R \label{GN2}
\eea
Here $v_F,v_H$ are the Fermi velocity of the 1DEG and the spinon velocity of the HC respectively. The double dots denote  normal ordering.

The model (\ref{model3}) was studied in \cite{Zachar}. 
 It has an emergent high SU(2)$_{charge}\times$SU(2)$_{spin}\times$SU(2)$_{spin}$ symmetry.  As we can see, the low energy Hamiltonian is split into three mutually commuting parts describing charge and spin excitations. The charge  excitations described by (\ref{charge}) are the same as in 1DEG. Their Hamiltonian does not contain any interacting terms and describes collective charge excitations of 1DEG (plasmons). In the absence of long range Coulomb interaction these plasmons have linear gapless spectrum $\omega = v_F|k|$.

 It is remarkable that the spin excitations are also separated in two independent sectors (\ref{GN1},\ref{GN2}) distinguished by parity. These two sectors are mirror images of each other.  At $J_K=0$ spin excitations of both 1DEG and spin S=1/2 Heisenberg chain are gapless spin density wave modes (spinons). Their spectra are $\omega_{1DEG} = v_F|k|$ and $\omega_{H} = v_H|k|$ respectively. As for 1D critical theories with linear spectrum, modes with different chirality do not interact with each other. So each spin sector consists of right- and left moving modes which do not talk to each other. They start talking once the interaction is turned on, but, as is clear from  (\ref{GN1},\ref{GN2}) it happens in a nontrivial way. Due to the incommensurability of the 1DEG and the Heisenberg chain the only components of the magnetization which interact are the smooth ones and they are expressed solely in terms of the currents. The only relevant interactions are those which include the currents of opposite chirality. As a consequence, the right movers from 1DEG interact with the left movers from the Heisenberg chain and vice versa. This is reflected in the structure of (\ref{GN1},\ref{GN2}) where we have grouped together the interacting modes.  So, as we have said,   the spin sector is decoupled  into  two independent sectors with different parity.

 As we have said, the charge subsector of the KH ladder  (\ref{charge}) is critical, the spectrum is linear: $\omega = v_F|k|$. Models (\ref{GN1},\ref{GN2}) describing the spin sector are integrable, at $J_K >0$ their spectrum consists of gapped spin 1/2 excitations (spinons) \cite{andrei} with dispersion relations $E(k)_{Lr} = E(-k)_{Rl} =E(k)$,
\bea
 E(k) = k(v_H-v_F)/2 +\sqrt{k^2(v_F+v_H)^2/4 + \Delta^2}, \label{disp}
\eea
where $
\Delta = C\sqrt{J_KJ_H}\exp[- \pi(v_F + v_H)/J_K], $ with $C$ being a nonuniversal numerical factor.
As is clear from (\ref{GN1},\ref{GN2}), the spinon gaps are generated by paring of spinons of a given chirality from the 1DEG with their partners of opposite chirality from the HC. This  makes the  ground state topologically nontrivial . For periodic boundary conditions (BC) the two spin sectors are independent and hence the energy levels are doubly degenerate.  For open BC the system can be projected onto one sector with periodic BC, but of the double length. The distinct feature of this spin liquid is that it can be formed only by conduction electrons and Heisenberg spinons acting {\it together}. In that respect our scenario radically differs from the one taken in \cite{senthil,subgeorges,paschen,pepin}, where the spin liquid can be formed solely by local magnetic moments.

 In one dimension continuous symmetry cannot be spontaneously broken even at zero temperature due to strong fluctuations. Hence one cannot have order, but there may be a tendency to it giving rise to a $T \rightarrow 0$ singularity in the corresponding static susceptibility. For example 1D Charge Density wave with wave vector $Q$ would have the static charge susceptibility at wave vector $Q$ diverging as $T^{-\alpha}$ with $\alpha >0$. The Fourier component of the charge density with wave vector $Q$ is then called fluctuating order parameter field or just order parameter field for short.

\subsection{Order parameters}

 The advantage of using the WZNW representation is that it makes the SU(2) symmetry manifest. However, for many practical calculations it is convenient to use the Abelian bosonization. This is possible since the SU$_1$(2) Kac-Moody algebra admits an Abelian representation. The corresponding Hamiltonian  can be written as the Hamiltonian of free bosons:
\bea
{\cal H}_{charge} = \frac{v_F}{2}\Big[(\p_x\Theta_c)^2 +(\p_x\Phi_c)^2\Big], \label{gauss}
\eea
where  field $\Phi_c$ and its dual field $\Theta_c$ satisfy the standard commutation relations $[\Phi_c(x),\p_x\Theta_c(x')] = \ri\delta(x-x')$. Likewise the Hamiltonians for the spin sector of the 1DEG and the S=1/2 Heisenberg chain can be written in the same Gaussian form with bosonic fields $\Phi_s,\Theta_s$ and $\Phi_H,\Theta_H$ respectively. The SU(2) symmetry imposed on the Gaussian model manifests itself in the selection of the operators constituting the operator basis of the theory (see below).  

 Model (\ref{gauss}) is critical, the excitation spectrum is linear. Hence its Hilbert space factorizes into holomorphic and antiholomorphic parts. This agrees with the fact that the WZNW Hamiltonian can be written as a sum of commuting parts containing currents of different chirality. In fact, the Gaussian model (\ref{gauss}) has a unique property among the critical models: its primary fields can be factorized into a product of holomorphic and antiholomorphic parts containing exponents of holomorphic $\varphi_a$ and antiholomorphic $\bar\varphi_a$ ($a= c,s, H$ parts of the bosonic fields 
\be
\varphi = (\Phi+\Theta)/2, ~~ \bar\varphi =   (\Phi-\Theta)/2.
\ee
For instance,  the bosonization rules for the fermion operators are 
\bea
R_{\s} = \frac{\xi_{\s}}{\sqrt{2\pi a_0}}\re^{-\ri\sqrt{2\pi}(\varphi_c + \s\varphi_s)}, ~~ L_{\s} = \frac{\xi_{\s}}{\sqrt{2\pi a_0}}\re^{\ri\sqrt{2\pi}(\bar\varphi_c + \s\bar\varphi_s)}, \label{fermions}
\eea
where $\xi_{\s}$ are Klein factors $\{\xi_{\s},\xi_{\s'}\} = 2\delta_{\s\s'}$. 

 The SU(2) symmetry manifests itself in the selection of the operators. The SU(2)-symmetric operator basis contains only derivatives of fields $\varphi, \bar\varphi$ and  integer powers of the  exponents 
\bea
&& z_{\s} = (2\pi a_0)^{-1/4}\exp[\ri\s\sqrt{2\pi}\varphi], ~~ \bar z_{\s} = (2\pi a_0)^{-1/4} \exp[-\ri\s\sqrt{2\pi}\bar\varphi], ~~ \s = \pm 1.\nonumber\\
&& z_{\s} = z^+_{-\s}. \label{z}
\eea
The chiral fields $z_{\s}^a, \bar z_{\s}^a$ have conformal dimensions (1/4,0), (0,1/4) respectively and can be considered as the holon and the spinon operators of the 1DEG ($a= c,s$)  and the spinon operators  of the Heisenberg chain ($a=H$). According to (\ref{fermions}) the annihilation operators of the right- and left moving electrons can be written as 
\bea
R_{\s} = \xi_{\s}\Big(z^c_{-}z^s_{\s}\Big), ~~ L_{\s} = \xi_{\s}\Big(\bar z^c_{-}\bar z^s_{\s}\Big). \label{RL}
\eea
The operator $\Delta_{cdw} = R^+_{\s}L_{\s} = (\bar z^c_+ z^c_-)(\bar z^s_{-\s}z^s_{\s})$. 

 The  S=1/2 Heisenberg model also possesses an approximate symmetry between correlation functions of the staggered components of the energy density and the magnetization operators such that they can be united in a single SU(2) matrix field
\bea
\hat G(x) = (-1)^n\Big[A({\bf S}_n{\bf S}_{n+1}) + \ri B(\vec\s{\bf S}_n)\Big], ~~ x= a_0n,
\eea
where $A,B$ are nonuniversal amplitudes. The symmetry is not perfect due to the marginally irrelevant current-current interaction (not shown here). This field is the spin 1/2 primary field of the SU$_1$(2) WZNW model. It can be factorized:
\bea
G_{\s\s'} = \frac{1}{\sqrt 2} \re^{\ri\pi(1-\s\s')/4}z^H_{\s}[\bar{z^H}_{\s'}]^+. \label{G}
\eea

Models (\ref{GN1},\ref{GN2})  have their own order parameters (OPs) with nonzero vacuum expectation values. 
\bea
\la {\cal O}_{rL}\ra = \sum_{\s}\la z^s_{\s}[\bar{z^H}_{\s}]^+\ra, ~~ \la {\cal O}_{lR}\ra = \sum_{\s}\la [{\bar z^s}_{\s}]^+ z^H_{\s}\ra. \label{ops}
\eea
They  form the amplitude of the composite OPs given by Eq. 9 of the main text:
\be
A = \la {\cal O}_{rL}\ra \la {\cal O}_{lR}^+\ra \label{A}
\ee
and are nonlocal in terms of both the 1DEG fermions and the local spins (hereafter they will be called  NOPs, with N for "nonlocal"). 
Since the scaling dimension of these NOPs is equal to 1/2, their vacuum expectation value $\sim \Delta^{1/2}$. To get a better understanding of the NOPs,  models (\ref{GN1},\ref{GN2}) will be rewritten using Abelian bosonization. For simplicity we will consider the case $v_H = v_F = 1$ when these models  are equivalent to the sine Gordon model with the Lagrangian:
\bea
L = \int \rd x \Big[\frac{ (1+ J_K/\pi )}{2}(\p_{\mu}\Phi)^2 - \frac{J_K}{\pi a_0}\cos(\sqrt{8\pi}\Phi)\Big],
\eea
where $\Phi = \varphi_s +\bar\varphi_H$ for the model (\ref{GN1}) or $\varphi_H + \bar\varphi_s$ for the model  (\ref{GN2}). The NOPs  (\ref{ops}) correspond to $\la \cos(\sqrt{2\pi}\Phi)\ra$. In the ground state this vacuum average may have any sign. Since only  the product (\ref{A}) enters into observable quantities, the ground state degeneracy is 2. This corresponds to the ground state degeneracy of spin S=1/2 antiferromagnetic chain. 

\subsection{Correlation functions}

According to  (\ref{RL}) the single particle Green's function factorizes into a product of two independent functions determined by the charge and the spin sector respectively. Thus for the right movers we have:
\bea
G_{RR} = \la\la z_-^c(\tau,x)[z^c_-]^+(0,0)\ra\ra \la\la z_{\s}^s(\tau,x)[z^s_{\s}]^+(0,0)\ra\ra, \label{GRR}
\eea
with a similar expression for the left movers with $z$ substituted for $\bar z$. 
Likewise, for the staggered magnetizations of the Heisenberg chain we have 
\bea
\la\la {\bf N}(\tau,x){\bf N}(0,0)\ra\ra = \la\la z^H(\tau,x)[z^H(0,0)]^+\ra\ra \la\la \bar z^H(\tau,x)[\bar z^H(0,0)]^+\ra\ra \label{NN}
\eea 

Since the charge sector is described by the Gaussian  noninteracting theory (\ref{gauss}), the corresponding correlator in (\ref{GRR}) is easy to calculate:
\bea
\la\la z_-^c(\tau,x)[z^c_-]^+(0,0)\ra\ra \sim (v_F\tau -\ri x)^{-1/2} \label{holon}
\eea
The next problem is to calculate the correlator of the spin components which enter in (\ref{GRR}) and (\ref{NN}). 
 As was explained in \cite{ZL,essler}, most of the spectral weight in these correlators  comes from the processes with an emission of a single massive spinon. Therefore it is sufficient to calculate just one matrix element. This was done using the Lorentz symmetry considerations \cite{essler}. Such considerations are directly applicable for the case $v_H =v_F$, but the general situation can be continuously deformed into the Lorentz invariant one.

These correlation functions can be calculated using the  advanced methods available in 1D. Some results can be obtained by  using a minimal knowledge of the spectrum and the operator structure of the theory, combined with symmetry considerations, as was done in \cite{essler,cdw}. In particular, the single electron Green's functions are similar to the ones in the Hubbard model and in the model of the 1DEG with attractive interaction  \cite{essler, mou}.
For small $|k| << k_F$ for the retarded functions we have $G(\omega, k\pm k_F) = G_{RR, LL}(\omega,k), ~~ G_{RR}(\omega,k) = G_{LL}(\omega,-k)$,
\bea
  && G_{RR}(\omega,k) =  \label{Green}\\
  && \frac{Z_0}{\omega -v_Fk}\Big[\frac{\Delta}{\sqrt{-(\omega +\ri 0 -v_Fk)(\omega +\ri 0 +v_Hk) + \Delta^2}}-1\Big] +...,\nonumber
\eea
where the dots stand for terms with emission of more than one spinon and $Z_0$ is a nonuniversal numerical factor.
At  $\omega =0$ the Green's function changes sign by going through  zero at wave vectors $\pm k_F$. Since the Green's function for the localized electrons has zeroes at $\pm \pi/2a_0$,  the total volume  inside of the surface of zeroes  is $\pi/a_0 + 2k_F$. It  includes both the localized and the delocalized electrons  in a full agreement with Luttinger theorem.

 Due to the decoupling of the spin sector  (\ref{GN1},\ref{GN2}) the correlators of the staggered parts of the magnetizations are products of the spinon Green's functions. The most singular parts of the dynamical magnetic susceptibilities are concentrated near the wave vectors $\pi/a_0$ (for the spins) and $2k_F$ (for the electrons). They have the following form \cite{Tsvelik2016}:
\bea
&& \la\la {\bf S}(\tau,x){\bf S(0,0)}\ra\ra = \label{chi}\\
&& \frac{Z_N (-1)^{x/a_0}}{\pi\sqrt{v_H^2\tau^2 + x^2}}\exp\Big\{-\Delta[(\tau + \ri x/v_H)(\tau - \ri x/v_F)]^{1/2}\Big\}\times \exp\Big\{- \Delta[(\tau - \ri x/v_H)(\tau + \ri x/v_F)]^{1/2}\Big\}, \nonumber\\
&& \la\la {\bf s}(\tau,x){\bf s}(0,0)\ra\ra = \la\la \rho(\tau,x)\rho(0,0)\ra\ra = \frac{Z_{s,\rho} \cos(2k_Fx)}{k_F\sqrt{v_F^2\tau^2 + x^2}}\la\la {\bf S}(\tau,x){\bf S(0,0)}\ra\ra, \label{chiDEG}
\eea
where $Z_N,Z_{s,\rho}$ are nonuniversal numerical factors. In the frequency -momentum space the susceptibility of the local spins (\ref{chi}) displays a strong continuum centered at $k = \pm \pi/a_0$ and the charge and spin susceptibilities of the 1DEG (\ref{chiDEG}) display a weaker continuum around $\pm 2k_F$. At $v_H=v_F =v$ the Fourier transform of (\ref{chi}) is
\be
\chi_{1D}(\omega, \pi/a_0 + k) = \frac{ Z_N}{\sqrt{4\Delta^2 + (vk)^2- \omega^2}} \label{chi1}
 \ee

\subsection{Coupled chains}

To promote the above OPs to the status of a real long range order we need to assemble the KH ladders into an array. To preserve the controllable status of the model we consider the layered array where the spin chains couple to spin chains and the electronic chains to the electronic ones.




For simplicity we consider only the nearest neighbor interactions and take into account the most relevant part of the exchange interaction:
\bea
&& H_{tunn} = \sum_{y,r} t_r\int \rd x [\psi^+_y(x)\psi_{y+r}(x) + H.c.],  \label{tunn}\\
&& H_{ex} = \sum_{y,r} \tilde J_r\int \rd x {\bf N}_y(x){\bf N}_{y+r}(x), \label{exchange}
\eea
where ${\bf N}$ is the staggered component of the magnetization. The exchange integral $\tilde J_r$ is a sum the direct antiferromagnetic  superexchange between the spin chains and the ferromagnetic one
   generated in the 2nd order in the interchain tunneling:
\bea
 && \tilde J = -t^2J_K^2\int \frac{\rd\omega\rd k}{(2\pi)^2}\frac{1}{[\ri\omega -\epsilon(k)]^2[\ri\omega -\epsilon(k+ \pi/a_0)]^2}  \nonumber\\
&& \sim - J_K^2t^2/W^3,
\eea
Hence $\tilde J$ may have any sign and magnitude.

The simplest situation  is the one when the interchain tunneling $t$ and exchange interaction $\tilde J$ are much smaller than the spin gap $\Delta$. 
However, in the present case we are interested in a different situation, namely, when the tunneling and the superexchange are sufficiently strong to create, respectively, a quasiparticle Fermi surface and a magnetic order. 

The electron Green's function can be in the first approximation obtained using RPA:
\bea
G(\omega, {\bf k}) = \Big[G^{-1}_{1D}(\omega, k_x) - \sum_r t_r\cos({\bf k r})\Big]^{-1}, \label{RPAG}
\eea
where $G_{1D}$ is given by (\ref{Green}). Since $G_{1D}(0,k_x)$ is finite, one needs the tunneling integral to exceed a certain threshold to create gapless quasiparticles. Hence below this threshold there is no Fermi surface and all single electron excitations are gapped. On the other hand, since $G_{1D}(0, \pm k_F) =0$, the Green's function (\ref{RPAG}) has zeroes at lines $k_x = \pm k_F$ at zero frequency and hence satisfies the Luttinger theorem, as we discussed above. 
The single electron Green's function (\ref{RPAG}) acquires poles at zero frequency when the interchain tunneling exceeds some critical value. These poles correspond to spinon-holon bound states which carry quantum numbers of electron and hence correspond to quasiparticles. Strictly speaking, for RPA to become a controllable approximation one needs a long range tunneling, something like $t_r \sim \exp(- |r|/R)$. Then corrections to RPA will include the small parameter $a_0/R$, where $a_0$ is the size of the unit cell. However, to simplify the discussion we will consider 2D rectangular lattice with nearest neighbor hopping.

  Obviously, RPA is approximation and one cannot guarantee that the lines of zeroes will not move when corrections are taken into account such that the sum rule  remains fulfilled, but (\ref{RPAG}) shows at least that the matter deserves to be taken seriously. The robustness of the RPA result with respect to corrections is discussed in \cite{Tsvelik2016}. First, the main threat to stability of RPA comes from singular corrections. All other corrections can be made parametrically small if we adopt long range tunneling. Second, there are two sources of singular corrections. There are those which are related to the proximity to the phase transition discussed in the previous subsection. However, since the transition temperature $T_c \sim {\cal J} = \sim {\tilde J}(t/\Delta)^2$ ( see the next Section) is parametrically different from the Fermi energy of the QP's, one can always find a parameter range where there exists a temperature window where the quasiparticles are well defined. Singular corrections can be also generated by interactions between the quasiparticles and collective modes. However, in the present case they are absent since the wave vectors of the order parameters ($2k_F + \pi/a_0$ and $\pi/a_0$) are incommensurate with $2k_F$ and although the Fermi surfaces of the QP's are nested, the order parameter fluctuations do not couple to them.

The model we are discussing describes  a compensated metal. The numbers of holes and electrons in the Fermi pockets are equal and the pockets do not contribute either to the Luttinger volume or to the Hall constant $R_H$. As before, the Luttinger theorem is fulfilled  solely by the Green's functions' zeroes. 

RPA expression for  the magnetic susceptibility is given by 
\bea
\chi(\omega, {\bf k}) = \Big[\chi^{-1}_{1D}(\omega, k_x) - \sum_r J_r\cos({\bf k r})\Big]^{-1}, \label{RPAchi}
\eea
where $\chi_{1D}$ is given by (\ref{chi}).  The latter formula is valid only for sufficiently small $J_r$ away from the instability. When the interstripe exchange exceeds the spin gap the system acquires a finite staggered magnetization. 

 The presence of static staggered moment or(and) of a quasiparticle Fermi surface leads to decrease in the order parameter amplitude. However, it seems reasonable that the pairing can survive if the ordered moment is small and the Fermi surface volume is much smaller than the volume of the bare Fermi surface. 
 
\subsection{Luttinger theorem}

One dimensional systems provide us with clear examples of states where spectral gaps develop when the Brillouin zone is not completely occupied by the electrons. Since Fermi surface in this case is absent, it is legitimate to ask what happens with the Luttinger theorem. It has been a matter of debate; in particular Dzyaloshinskii remained us that in the standard proof of the Luttinger theorem zeroes of the single particle Green's function contribute to the Luttinger volume alongside with the poles \cite{IgorD}. This suggestion has been challenged, the most persuasively in \cite{kane}, where it was argued that the standard proof cannot be used in the corresponding models since the  Luttinger-Ward functional does not exist if the self energy has poles. However, we know the gapped systems where  the fulfillment of the Luttinger theorem can be demonstrated explicitly, as it was done in \cite{mybook}. These include all (1+1)-dimensional models with Lorentz symmetry where the Green's functions of the right- and left moving particles situated at $+k_F$ and $-k_F$ respectively are expressed as 
\be
G_{RR,LL}(\tau,x) = m\re^{\pm \ri\phi}{\cal F}(m r), ~~ r= \sqrt{\tau^2 + (x/v)^2}, ~~ \phi = \tan^{-1}(\tau v/x), 
\ee
where $m$ is the characteristic energy scale of order of the gap. The function ${\cal F}(x)$ decays exponentially at large $x$ and $\sim 1/x$ at $x<<1$. Then we have 
\bea
G_{RR,LL} (\omega =0, k=0) \sim \int_0^{2\pi}\rd\phi\int_0^{\infty}\rd r r{\cal F}(mr) =0.
\eea
The integration can be performed safely since the integral in $r$ converges. The proof can be easily extended for the cases when charge and spin sectors have different velocities. This shows that at least  for model (\ref{model3}) the Luttinger theorem with zeroes is valid. 

\subsection{Ginzburg-Landau action for collective modes}

 In what follows it will be assumed that the coupling between KH ladders is unfrustrated and the  most of the spectral weight in  the   spin sector lies at energies higher than the spin gap. The latter  is equivalent to the assumption that the quasiparticle Fermi surface and the staggered magnetic moment are small.  Then  the spin sector can be effectively integrated out and one obtains the effective action for the low energy charge modes. They are bosons. There are neutral and  charged ones, the latter ones carry  charge $2e$. In the  dimerized phase the order combines dimerization in the spin subsystem and the conventional superconductivity in the electron one.

 At  certain temperature this fluctuating superconductor undergoes a phase transition into a state with either an odd-frequency pair density  or odd-frequency charge density wave order.   The effective Ginzburg-Landau action for the collective modes is written in terms of the SU(2) matrix field $g$ (Eq. 9 from the main text) describing the low energy components of the composite OPs. The partition function is the path integral $Z = \int Dg \exp(-S)$   with action
\bea
  S = \sum_y\Big[ W[g_y] -   \int_0^{1/T} \rd\tau\int \rd x {\cal J}_r\mbox{Tr}(\s^zg_y\s^zg^+_{y+r} +H.c.)\Big], \label{GL}
\eea
where $W[g]$ is the SU$_1$(2) WZNW action and  ${\cal J} \sim {\tilde J}(t/\Delta)^2$. In the Hamiltonian formulation $W[g]$ corresponds to (\ref{charge}). Notice that the sign of ${\cal J}$ coincides with the sign of the interchain exchange interaction.  For the sake of simplicity the above formulae are written for a bipartite lattice with nearest neighbor coupling. 
In the presence of long range Coulomb interaction this action must be augmented by the term including smooth parts of the charge density.
In non-Abelian bosonization 
\bea
J_R = -\frac{k}{4\pi}g(\p_{\tau} +\ri\p_x)g^{-1}, ~~ J_L= -\frac{k}{4\pi}g(-\p_{\tau} +\ri\p_x)g^{-1}
\eea
Hence
\bea
\rho = (J_L + J_R)^z = -\frac{\ri}{2\pi}\mbox{Tr}(\s^z g\p_x g^{-1}),
\eea
and the Coulomb interaction yields
\bea
V = \frac{1}{2}\sum_{y,y'}\int \rd x\rd x' \rho(x,y)U(x-x';y-y')\rho(x',y'). \label{coulomb}
\eea
Although the Wess-Zumino term in $W[g]$ which includes time derivative does not change the critical properties of the transition at finite T, its presence affects the dynamics of the charge excitations.  Above the transition temperature the correlation function of the OP's can be estimated using RPA:

\bea
&& \la\la g(\omega,{\bf k})g^+(\omega,{\bf k})\ra\ra = \Big[D^{-1}(\omega,k_{\parallel}) - \sum_{\bf r} {\cal J}_r \cos({\bf k r})\Big]^{-1}, \label{transition}\\
 && D(\omega,k) = \frac{A}{T}\rho\Big(\frac{\omega - v_Fk}{4\pi T}\Big)\rho\Big(\frac{\omega + v_Fk}{4\pi T}\Big), ~~ \rho(x) = \frac{\Gamma(2\Delta - \ri x)}{\Gamma(1-2\Delta -\ri x)}.
\eea
Here $D$ is the correlation function in the SU$_1$(2) WZNW model, the sum runs over nearest neighbors of a given ladder and $A$ is a nonuniversal amplitude. Eq.(\ref{transition}) yields estimate of the mean field transition temperature: $T_{MF} \sim {\cal J}$.   If the lattice is three dimensional the real and mean field transition temperatures are not that different from each other.

It is instructive to calculate the contribution of the bosonic modes to the transverse conductivity above the transition. In the leading order in ${\cal J}$ is given by
\bea
\s_{\perp}(\omega) \sim \frac{{\cal J}^2}{T}\sin(4\pi\Delta)\frac{1}{\omega}\Im m \frac{\Gamma^2(2\Delta - \ri\omega/4\pi T)}{\Gamma^2(1-2\Delta - \ri\omega/4\pi T)}
\eea
Here we need to take into account the fact that $\Delta = 1/4K_c$, where $K_c$ is the Luttinger parameter which we have  treated so far as equal to 1.
In the low frequency limit we have
\bea
\s_{\perp}(0) \sim \frac{{\cal J}^2}{T^2} \cos^2(\pi/2K_c).
\eea

Below the transition we can extract from (\ref{GL}) the Ginzburg-Landau free energy by neglecting the time dependency in $W[g]$:
\bea
F = \sum_y \int \rd x \Big[\frac{v_F}{16\pi}\mbox{Tr}(\p_x g_y^+\p_x g_y) - \frac{1}{2}\mbox{Tr}\sum_r {\cal J}_r(\s^zg_y\s^zg^+_{y+r} +H.c.)\Big], \label{F}
\eea
which can also be augmented by the Coulomb interaction term (\ref{coulomb}). Then taking the continuum limit in the direction perpendicular to the chains we can rewrite  (\ref{F}) as follows:
\bea
&& F = \frac{1}{2}\int \rd V \Big(\omega^3_{\mu}\rho^{\parallel}_{\mu\nu}\omega^3_{\nu}  + \p_{\mu}{\bf n}\rho^{\perp}_{\mu\nu}\p_{\nu}{\bf n}\Big), \\
&& \omega^3_{\mu} = \p_{\mu}\phi  - \cos\theta\p_{\mu}\psi, ~~ {\bf n} = (\cos\theta, \sin\theta\cos\psi, \sin\theta\sin\psi).
\eea
This free energy is similar to the one of He$^3$-A.

  To include the magnetic field we have to take into account  that some components of the $g$-matrix are charged fields with charge $2e$.  As is clear from the main text, the charged component of the order parameter is the  off-diagonal part of $g$. Therefore under gauge transformation $g$ transforms as $\re^{\ri\alpha}g\re^{-\ri\alpha}$. It will be convenient to us to choose the Euler parametrization for $g$:
\bea
g = \exp\Big(\frac{\ri}{2}\s^z\phi\Big)\exp\Big(\frac{\ri}{2}\s^x\theta\Big)\exp\Big(\frac{\ri}{2}\s^z\psi\Big).
\eea
The gauge transformation shifts $\phi$ and $\psi$ in opposite directions. In the continuum limit we have to replace:
\bea
\p_{\mu}\phi  \rightarrow  \p_{\mu}\phi - (2e/c)A_{\mu}, ~~ \p_{\mu}\psi \rightarrow  \p_{\mu}\psi + (2e/c)A_{\mu}.
\eea

 \end{widetext}

\end{document}